\documentstyle[12 pt]{article}

\newcommand{\be}{\begin{equation}}
\newcommand{\ee}{\end{equation}}
\newcommand{\teta}{\theta}
\newcommand{\p}{\partial}
\newcommand{\beqn}{\begin{eqnarray}}
\newcommand{\eeqn}{\end{eqnarray}}
\newcommand{\nn}{\nonumber}

\newcommand{\f}{\frac}

\newcommand{\bfl}{\begin{flushleft}}
\newcommand{\efl}{\end{flushleft}}
\newcommand{\br}{\begin{flushright}}
\newcommand{\er}{\begin{flushright}}
\newcommand{\LDBI}{\cal L_{DBI}}

\begin{document}
\begin{titlepage}
\flushright{IP/BBSR/2002-14}
\flushright{hep-th/0205303}

\vspace{1in}

\begin{center}
\Large
{\bf Solution  to  worldvolume action of D3 brane in pp-wave background}
\vspace{1in}

\normalsize

\large{ Shesansu Sekhar  Pal }\\
{e-mail: shesansu@iopb.res.in}

\normalsize
\vspace{.7in}

{\em Institute of Physics \\
Bhubaneswar - 751005, India }

\end{center}

\vspace{1in}

\def    \beq    {\begin{equation}}

\baselineskip=24pt
\begin{abstract}
We find  nonsupersymmetric and supersymmetric solutions
 of D3 brane configuration 
in the background of pp wave obtained as a Penrose 
limit of $AdS_5\times S^5$.

\end{abstract}

\end{titlepage}

Recently, it has been shown that the study of  maximally 
supersymmetric IIB supergravity
background, pp waves\cite{blau}, of various geometries gives rise to many
interesting ideas in the context of AdS/CFT correspondence. Especially, 
the study of BMN\cite{bmn} tells us how to see the duality in a
particular sector of the four dimensional SU(N) gauge theory with
${\cal N}=4$ with type IIB superstring theory in pp wave background. 
The study of  duality\footnote{Also studied in\cite{bgmnn}, see also
  refs. \cite{kpss}-\cite{kklp} for related recent developments.}  
have been extended to  
 various singular spaces with different supersymmetries in 
\cite{ikm}-\cite{ot} and have been studied in \cite{dgr}, \cite{lor}, 
\cite{kp} from holographic point of view. 
Moreover, string theory in this background is exactly solvable. 
The quantization of strings in the pp wave background with NS-NS and 
R-R fluxes is done in \cite{rm}-\cite{ft} and the quantization of Dp
brane is studied in \cite{dp}. Solution to branes in the pp wave
background have been studied in \cite{db}-\cite{st} and in \cite{bmz},\cite{ak}
from supergravity point of view. The authors of
\cite{st} have studied various configurations of Dp brane with
different embeddings and found solutions of constant  embeddings with  
zero worldvolume flux on the brane.

In this letter we shall derive nonsupersymmetric and supersymmetric 
solutions of D3
brane whose worldvolume directions are extended along  
any  one of the SO(4) directions and one along positive lightcone
direction and others are along one of the SO(4) directions 
in the pp wave background.

Before we start to present the solution of D3 brane in the pp wave 
background, let us discuss the  geometry briefly. 
The geometry of $AdS_p\times S^q$ in global coordinate can be
described as:
\be 
ds^2=R^2_A\big(d\rho^2+\sinh^2\rho
d\Omega^2_{p-2}-\cosh^2dt^2\big)+R^2_S\left(d\teta^2+\sin^2\teta
  d\Omega^2_{q-2}+\cos^2\teta d\psi^2\right),
\ee  
where $R_A, R_S$ are the radius of curvature and radius of  $AdS_p$
and $S^q$ respectively. To derive the pp wave, we have to consider a
particle moving along  the $\psi$ direction and sitting at $\rho=0$ 
and $\teta=0$ and the geometry seen by the particle while moving along 
this trajectory will give us the desired geometry. To do so, let us
define coordinates as:
\be
x^+=\f{1}{2}(t+\f{R_S}{R_A}\psi),\quad
x^-=\f{R^2_A}{2}(t-\f{R_S}{R_A}\psi),\quad x=R_A\rho,\quad
y=R_S\teta.
\ee
Hence, the geometry in this coordinates becomes
\beqn
ds^2&=&R^2_A\Bigg[\f{dx^2}{R^2_A}+\sinh^2\left(\f{x}{R_A}\right)d\Omega^2_{p-2}-\cosh^2\left(\f{x}{R_A}\right)\Bigg\{(dx^+)^2+R^{-4}_A(dx^-)^2+2R^{-2}_Adx^+dx^-\Bigg\}\Bigg]\nn \\
& &+R^2_S\Bigg[\f{dy^2}{R^2_S}+\sin^2\left(\f{y}{R_S}\right)d\Omega^2_{q-2}+\cos^2\left(\f{y}{R_S}\right)\Bigg\{\f{R^2_A(dx^+)^2}{R^2_S}+\f{(dx^-)^2}{R^2_SR^2_A}-\f{2}{R^2_S}dx^+dx^-\Bigg\}\Bigg]
\eeqn 
Let us define a limit
\be
R_A\rightarrow\infty,\quad R_S\rightarrow\infty,\quad{\rm and\quad  keeping\quad \f{R_S}{R_A}=fixed}.
\ee 
In this limit the metric becomes 
\be
ds^2=-4dx^+dx^--\left(x^2+\f{R^2_A}{R^2_S}y^2\right)(dx^+)^2+dx^2+x^2d\Omega^2_{p-2}+dy^2+y^2d\Omega^2_{q-2},
\ee
scaling $x^+\rightarrow \mu x^+$ and $x^-\rightarrow\f{x^-}{\mu}$, we
get the metric as 
\be 
ds^2=-4dx^+dx^--\mu^2\left(x^2+\f{R^2_A}{R^2_S}y^2\right)(dx^+)^2+dx^2+x^2d\Omega^2_{p-2}+dy^2+y^2d\Omega^2_{q-2}.
\ee 
It's easy to see from the Penrose limit of  $AdS_p\times S^q$ that the 
Penrose limit of  spaces with different values of p and q but with
same  p+q have the same pp wave provided the radius of curvature of
AdS is equal to the radius of sphere. For $AdS_5\times S^5$ the
corresponding R-R 5-form field strength is 
\be
F_{+1234}=F_{+5678}={\rm constant}\times \mu.
\ee
The world volume theory of Dp brane supports solitonic configurations
of lower dimensional branes. In flat space, in particular, in the
context of D3 branes, BPS monopoles correspond to an orthogonal D1
brane. We shall describe D3 brane in the Penrose limit of 
 $AdS_5\times S^5$ background.

The low energy dynamics of a D3 brane in a pp wave background, i.e
\be
ds^2=-4dx^+dx^--\mu^2(x^2_1+\ldots+x^2_8)(dx^+)^2+\sum^8_{i, j=1}dx^idx^i\eta_{ij}, 
\ee
where we have taken the radius of curvature of AdS is same as the
radius of the sphere, 
is described by 
\be
\label{DBI_action}
S=-T_3\int d^4\sigma \sqrt{|det(P[G]_{ab}+\lambda F_{ab})|},
\ee
where we have set $B_{\mu\nu}=0$, dilaton=0.  
$F_{ab}$ is the U(1) field strength living on the brane, 
$\lambda\equiv 2\pi\alpha^{'}$, but we shall set $\lambda=1$, henceforth,
and P is the pullback which will pullback the 
bulk fields onto the worldvolume of the brane and $P[G]_{ab}$ is given by 
\be
P[G]_{ab}=-2\p_aX^+\p_bX^--2\p_aX^-\p_bX^+-\mu^2(x^2_1+\ldots+x^2_8)\p_aX^+\p_bX^++\sum^8_{i, 
  j=1}\p_aX^i\p_bX^j\eta_{ij}.
\ee
We shall choose the static gauge choice as\footnote{In our notation
  a,b denotes the worldvolume coordinates of D3 brane and can take
  values 1,2,3,4 and  m can take values from 5 to 8.} 
\be
X^+=\phi^9,\quad X^-=\phi^{10},\quad X^1=\sigma^1,\quad X^2=\sigma^2,\quad
X^3=\sigma^3,\quad X^4=\sigma^4,\quad X^m=\phi^m.
\ee 
In this choice of static gauge  pullback becomes
\be
\label{pullback}
P[G]_{ab}=+\p_a\sigma^1\p_b\sigma^1+\p_a\sigma^2\p_b\sigma^2+\p_a\sigma^3\p_b\sigma^3+\p_a\sigma^4\p_b\sigma^4+\p_a\phi\p_b\phi,
\ee
where we have excited only one transverse
scalar, say $\phi=\phi^5$. Let us also excite a magnetic field 
 $B^{\alpha}=\f{1}{2}\epsilon^{\alpha\beta\gamma}F_{\beta\gamma}
(\alpha,\beta,\gamma=2,3,4)$, and  shall take the 'time' as $\sigma^1$. 
For static configurations, the energy
then becomes
\beqn
\label{energy}
E=-L&=&T_3\int d^3
\sigma\Bigg[(1\pm\vec{B}\cdot\vec{\nabla}\phi)^2+(\vec{B}\mp\vec{\nabla}\phi)^2\Bigg]^{\f{1}{2}}\nn \\
& &T_3\geq\int d^3\sigma(1\pm\vec{B}\cdot\vec{\nabla}\phi)
\eeqn

The lower bound is achieved when 
\be 
\label{b=p_phi}
\vec{B}=\pm\vec{\nabla}\phi,
\ee
using this along with the Bianchi identity,  $\vec{\nabla}\cdot\vec{B}=0$, 
we get the equation to
scalar as
\be
\label{equation_phi}
\nabla^2\phi=0.
\ee
The trivial solutions to  eq.(\ref{equation_phi}) and  magnetic field is 
\be
\label{solution}
\phi={\rm constant},\quad \vec{B}=0,
\ee
and the nontrivial solution is 
\be 
\label{solution_1}
\phi=\f{N}{2r},\quad~~~~~~~\vec{B}=\mp\f{N}{2r^3}\vec{r},
\ee
where $r^2=(\sigma^1)^2+(\sigma^2)^2+(\sigma^3)^2$ and N is an integer 
due to charge quantization \cite{cmt}. 
Hence, the energy to this configuration is 
the sum of energy of Born-Infeld  and Chern-Simon part. Since, we are
taking  minimum energy for the nonlinear Born-Infeld action implies
the total energy is minimum.  

In order to see how much supersymmetry does this configuration preserve 
let us note that 
the supersymmetric analysis to D3 brane have been studied 
in \cite{bp},\cite{ch},\cite{bmz} 
and in \cite{dp}, these 
 authors  have  noted that  D3 brane preserves half of
the supersymmetry in the background of pp waves provided the
worldvolume directions of D3 are along the light cone directions along
with two other directions are along  either of SO(4) directions. 
If one of the direction
is along one SO(4) and  other is along the other SO(4) direction then
the solution do not preserve any supersymmetry \cite{st}.

It has been argued that with every brane embedding there is a kappa
symmetry projection operator which  satisfies 
\be
\label{susy_condition}
\Gamma\epsilon=\epsilon,
\ee
if the given brane embedding preserves some fraction of the
supersymmetry, 
where $\epsilon$ is a Killing spinor of the supersymmetric background
and $\Gamma$ is kappa symmetric
projector and is defined as in \cite{cgnw}-\cite{bt} 
\be
\label{projector_in_general}
d^{p+1}\sigma\Gamma=-\f{e^{\f{p-3}{4}\phi}}{\LDBI} exp({ e^{-\f{\phi}{2}{\cal
    F}}})\wedge XY|_{vol},
\ee 
 with
\be 
X=\bigoplus_n \gamma_{(2n+q)} P^{n+q},
\ee
where
\beqn
& &IIA: P=\gamma_{11}\quad Y=1,\quad q=1\nn \\& &
IIB: P=K,\quad Y=I,\quad q=0,
\eeqn
where $|_{vol}$ means that term should be proportional to the volume
form of the brane and ${\cal F}=F-B$. The operators I and K act on spinors as
$I\psi=-i\psi$ and $K\psi=\psi^*$ respectively. These operators are
anticommuting and satisfy the following properties:  
$I^2=-1$, $J^2=1=K^2$ and IJ=K.
$\f{1}{\LDBI}$ is the value of the DBI Lagrangian evaluated in the
background with the embeddings. $\gamma_{(2n)}$ is a 2n-form 
gamma matrices defined on
target space as 
\be
\gamma_{(2n)}=\f{1}{2n!} \gamma_{i_1\ldots i_{2n}}
d\sigma^{i_1}\wedge\ldots\wedge d\sigma^{i_{2n}},
\ee  
and $\gamma_{i_1\ldots i_{2n}}$ is the pullback of the target space
gamma matrices 
\be
\gamma_{i_1\ldots i_{2n}}=\p_{i_1}X^{\mu_1}\ldots\p_{i_{2n}}X^{\mu_{2n}}\gamma_{\mu_1\ldots\mu_{2n}}.
\ee 
It has been shown in \cite{cgnw} that $\Gamma$ satisfies the following
properties. $\Gamma^2=1$ and tr$(\Gamma)=0$, which enables to define
projector of that kind which projects out half of the worldvolume 
fermions thus
making the degrees of freedom of worldvolume fermions and bosons same.

Since, we are studying a D3 brane configuration where dilaton and 2-form
antisymmetric B field is zero in a IIB background then the projector
written in 
eq. (\ref{projector_in_general}) becomes 
\be
\label{projector_1}
\Gamma=-\f{\epsilon^{i_1\ldots
    i_4}}{\LDBI}\left[\f{1}{24}\gamma_{i_1\ldots i_4}
    I+\f{1}{4}F_{i_1i_2}\gamma_{i_3i_4}J+\f{1}{8}F_{i_1i_2}F_{i_3i_4} I\right].
\ee 
Evaluating the projector in our static gauge choice, we get:
\beqn
\label{projector}
\Gamma&=&-\f{i}{\LDBI}\bigg[(\gamma_{1234}+\p_2\phi\gamma_{1345}-\p_3\phi\gamma_{1245}+\p_4\phi_{1235})I\nn 
\\& &+\big\{B^2(\gamma_{12}+\p_2\phi\gamma_{15})+B^3(\gamma_{13}+\p_3\phi\gamma_{15})+B^4(\gamma_{14}+\p_4\phi\gamma_{15})\big\}J\bigg].
\eeqn  
The first and second line of eq. (\ref{projector}) follows from the 
first and second term of eq. (\ref{projector_1}) and the last term
vanishes because of our choice of magnetic field. For negative
chirality type IIB spinors 
\be 
\gamma_{+-12345678}\epsilon=-\epsilon,
\ee
the Killing spinor, $\epsilon$,  is 
 derived by setting the supersymmetric variation 
of dilatino and gravitino to zero, of the pp-wave is given by
\cite{blau}, \cite{st}  
\beqn 
\label{Killing_spinor}
\epsilon&=&\bigg\{1-\f{i}{2}\sum^4_{a=1}\gamma_-(y^a\gamma_a\gamma_{1234}+z^a\gamma_{(a+4)}\gamma_{5678}\bigg\}\nn 
\\& &\bigg(\cos\f{x^+}{2}-i\sin\f{x^+}{2}\gamma_{1234}\bigg)\bigg(\cos\f{x^+}{2}-i\sin\f{x^+}{2}\gamma_{5678}\bigg)(\lambda+i\eta),
\eeqn
where $\lambda$ and $\eta$ are constant, real negative chiral spinors
and $y^a$, and $z^a$ are the coordinates along the first and second SO(4) 
respectively. 
 Substituting the value of $\Gamma$ and $\epsilon$ in
eq. (\ref{susy_condition}), and restricting to $\phi=\phi(\sigma^2)$
and only one component of magnetic field i.e. to $B^2$, for
simplicity, we get
\be
\label{condition} 
\epsilon=-\f{1}{\LDBI}\bigg[(\gamma_{1234}+\p_2\phi\gamma_{1345})\epsilon-B^2(\gamma_{12}+\p_2\phi\gamma_{15})\epsilon^*\bigg].
\ee 
Let us define 
\beqn
&P&\equiv-\f{1}{2}\sum^4_{a=1}\gamma_-(y^a\gamma_a\gamma_{1234}+z^a\gamma_{(a+4)}\gamma_{5678})\nn \\
&Q&\equiv B^2(\gamma_{12}+\p_2\phi\gamma_{15})\nn \\
&R&\equiv \gamma_{1234}+\p_2\phi\gamma_{1345}.
\eeqn
using these, we can rewrite eq. (\ref{condition}) as 
\be 
\label{condition_1}
\epsilon=-\f{1}{\LDBI}\bigg[ R\epsilon-Q\epsilon^*\bigg].
\ee
Since the Killing spinor eq. (\ref{Killing_spinor}) holds for all
values of $x^+$, implies, on plugging  into
eq. (\ref{condition_1}) gives rise to
\beqn
\label{condition_2}
& &\bigg(1+\f{R}{\LDBI}\bigg)(1+iP)(\lambda+i\eta)=\f{Q}{\LDBI} (1-iP)(\lambda-i\eta)\nn 
\\& &\bigg(1+\f{R}{\LDBI}\bigg)(1+iP)(\gamma_{1234}+\gamma_{5678})(\lambda+i\eta)=-\f{Q}{\LDBI}(1-iP)(\gamma_{1234}+\gamma_{5678})(\lambda-i\eta)\nn 
\\& &\bigg(1+\f{R}{\LDBI}\bigg)(1+iP)\gamma_{1\cdots
  8}(\lambda+i\eta)=\f{Q}{\LDBI}(1-iP)\gamma_{1\ldots 8}(\lambda-i\eta).
\eeqn

Equating  real and imaginary parts  of eq. (\ref{condition_2}) and
using negative chirality of $\lambda$ and $\eta$, we get
\beqn 
\label{condition_3}
&
&\bigg(1+\f{R-Q}{\LDBI}\bigg)(\lambda-P\eta)=0,\nn 
\\& &
\bigg(1+\f{R+Q}{\LDBI}\bigg)(P\lambda+\eta)=0,\nn \\& &
\gamma_-\gamma_+\bigg(1+\f{R-Q}{\LDBI}\bigg)\lambda=0,\nn \\& &
\gamma_-\gamma_+\bigg(1+\f{R+Q}{\LDBI}\bigg)\eta=0,\nn \\& &
\gamma_-\gamma_+\bigg(1+\f{R+Q}{\LDBI}\bigg)\gamma_{1234}\lambda=0,\nn \\&
&
\gamma_-\gamma_+\bigg(1+\f{R-Q}{\LDBI}\bigg)\gamma_{1234}\eta=0.
\eeqn
Let us see how much supersymmetry does this configuration preserves  
for zero magnetic field i.e. $B^2=0$. From  third and fourth equation
of eq. (\ref{condition_3}) we get 
\beqn
\label{con_l_e_1} 
& &(1+\gamma_{1234})\lambda=0,\nn \\& &
(1+\gamma_{1234})\eta=0,
\eeqn   
which also follows from fifth and sixth equation of
eq. (\ref{condition_3}). However, from first and second equation of
eq. (\ref{condition_3}) we get on invoking eq. (\ref{con_l_e_1})
\beqn 
\label{con_l_e_2}
& &(1-\gamma_{1234})\lambda=0\nn \\& &
(1-\gamma_{1234})\eta=0.
\eeqn
Compatibility of eq. (\ref{con_l_e_1}) and eq. (\ref{con_l_e_2})
implies that for zero magnetic field the configuration breaks all
supersymmetry. Doing the same kind of analysis for nonzero magnetic
field we conclude that the configuration  breaks all supersymmetry.

Let us look at a different configuration, (+,3,0), which means that
the worldvolume directions of the brane are extended along the
positive lightcone and along one of the SO(4) directions. In this case 
we shall choose our static gauge choice as
\be
X^+=\tau,\quad X^-=\phi^9,\quad X^1=\sigma^1,\quad X^2=\sigma^2,\quad
X^3=\sigma^3,\quad X^4=\phi^4,\quad X^m=\phi^m.
\ee 
In this choice of static gauge  pullback becomes\footnote{Here a,b can 
  take values 0,1,2,3}
\beqn
\label{pullback_1}
P[G]_{ab}&=&-\mu^2\left((\sigma^1)^2+(\sigma^2)^2+(\sigma^3)^2+\phi^4\phi^4+\phi^m\phi^m\right)\p_a\tau\p_b\tau+\p_a\sigma^1\p_b\sigma^1+\nn 
\\& &
\p_a\sigma^2\p_b\sigma^2+\p_a\sigma^3\p_b\sigma^3 
+\p_a\phi^4\p_b\phi^4+\p_a\phi^m\p_b\phi^m,
\eeqn
where we have taken the spacetime coordinates as same as the
embeddings, i.e. $x^i=X^i,~
i=1,\ldots,8$. Let us excite only one transverse
scalar, say $\phi=\phi^4$ and a magnetic field
$B^{\alpha}=\f{1}{2}\epsilon^{\alpha\beta\gamma}F_{\beta\gamma}
(\alpha,\beta,\gamma=1,2,3)$. For static configurations, the energy
then becomes
\beqn
\label{energy_1}
E=-L&=&T_3\int
d^3\sigma\sqrt{\mu^2\left((\sigma^1)^2+(\sigma^2)^2+(\sigma^3)^2+(\phi)^2\right)}\Bigg[(1\pm\vec{B}.\vec{\nabla}\phi)^2+(\vec{B}\mp\vec{\nabla}\phi)^2\Bigg]^{\f{1}{2}}\nn \\
& &T_3\geq\int d^3\sigma\sqrt{\mu^2\left((\sigma^1)^2+(\sigma^2)^2+(\sigma^3)^2+(\phi)^2\right)}(1\pm\vec{B}.\vec{\nabla}\phi)
\eeqn
The lower bound to energy and the solution are written in
eq. (\ref{b=p_phi}) and eq. (\ref{solution_1}). However, the
consistency of the static gauge choice implies that we have to find
solution to $\phi$ and B by taking the following restriction into account:
\be 
\mu^2\left((\sigma^1)^2+(\sigma^2)^2+(\sigma^3)^2+(\phi)^2\right)={\rm
  constant}\equiv c^2.
\ee
So, the solution to the restriction and the equation satisfied by B
and $\phi$ gives us
\be 
\phi={\rm constant},\quad B=0,\quad
(\sigma^1)^2+(\sigma^2)^2+(\sigma^3)^2={\rm constant}\equiv c^2_1,
\ee
by looking at the solution we conclude that the brane is a spherical
D3 brane. To check the supersymmetry preserved by this configuration,
let us take the projector as 
\be 
\Gamma=-\f{1}{c} (\gamma_{+}-\f{c^2}{2}\gamma_-)\gamma_{123}I.
\ee
Substituting the above form of $\Gamma$ and $\epsilon$ from
eq. (\ref{Killing_spinor}) in eq. (\ref{susy_condition}), we get 
the following equations.
\beqn
\label{con}
& &(1-i(Q-R))(1+iP)(\lambda+i\eta)=0\nn \\& &
(1-i(Q-R))(1+iP)\gamma_{1\ldots 8}(\lambda+i\eta)=0\nn \\& &
(1-i(Q-R))(1+iP)(\gamma_{1234}+\gamma_{5678})(\lambda+i\eta)=0,
\eeqn
where $Q=\f{\gamma_{+123}}{c}$ and $R=\f{c}{2}\gamma_{-123}$. Using
$\gamma^2_-=0$, we get from second and third equation of
eq. (\ref{con})
\be 
\gamma_+\lambda=0,\quad \gamma_+\eta=0.
\ee
On invoking this condition on $\lambda$ and $\eta$ in the first
equation of eq. (\ref{con}), we see that the configuration preserves
one half supersymmetry for $P=-\f{c}{2}\gamma_{-123}$. However, for 
$P=-\f{c}{2}\gamma_{-234}$, we get one more condition on $\lambda$ and 
$\eta$. Hence, for the later choice of P we conclude that the
configuration preserves one quarter supersymmetry. So, finally, for
different choices of P the configuration preserves different amount of 
supersymmetry.

We have derived nonsupersymmetric and supersymmetric solutions 
of D3 brane in the pp 
wave background for two different kind of configurations, in one case
the brane is extended completely along one of the SO(4) directions and in the
other case one of the worldvolume direction is extended along the
positive lightcone direction and the rest three are along  one of
the SO(4) directions. In the former case we found that our solution
breaks all supersymmetry for  zero magnetic field 
on the worldvolume of the brane and for a constant transverse
scalar and also for nonzero magnetic field and nonconstant transverse
scalar. Where as in the later configuration it preserves different
amount of supersymmetry depending on the  choices of P.

{\bf Acknowledgements}\\

I would like to thank Alok kumar for many useful discussions,  
Sunil Mukhi for many useful correspondences and to the anonymous
referee for several constructive suggestions.


\vspace{.2in}
\begin{center}
{\bf References}
\end{center} 
\begin{enumerate}
\bibitem{blau}
J. Figueroa-O'Farrill, G. Papadopoulos, 
``Homogeneous Fluxes, Branes and a Maximally Supersymmetric Solution of 
M-theory,'' {\it JHEP} {\bf 0108} (2001)  036; hep-th/0105308.
M. Blau, J. Figueroa-O'Farrill, C. Hull, G. Papadopoulos, 
``Penrose Limits and Maximal Supersymmetry,'' hep-th/0210081.
M. Blau, J. Figueroa-O'Farrill, C. Hull, G. Papadopoulos, 
``A New Maximally Supersymmetric Background of IIB Superstring Theory,''
{\em JHEP} {\bf 0201} (2002) 047, hep-th/0110242; 
M. Blau, J. Figueroa-O'Farrill, G. Papadopoulos,
``Penrose limits, supergravity and brane dynamics,''hep-th/0202111. 
\bibitem{bmn} D. Berenstein, J. Maldacena and H. Nastase, ``Strings in 
  flat space and pp waves from ${\cal N}=4$ super Yang Mills,''
  hep-th/0202021.
\bibitem{bgmnn} D. Berenstein, E. Gava, J. Maldacena, K.S. Narain and 
H. Nastase, ``Open strings on plane waves and their Yang-Mills duals,''
hep-th/0203249.
\bibitem{kpss} C. Kristjansen, J. Plefka, G.W. Semenoff and M. Staudacher, ``A
New double-scaling limit of N=4 super Yang-Mills theory and
pp-wave strings,''  hep-th/0205033.
\bibitem{gkp}
S.S. Gubser, I.R. Klebanov and A.M. Polyakov, ``A semi-classical
limit of the gauge/string correspondence,'' hep-th/0204051.
\bibitem{bn}
D. Berenstein and H. Nastase, ``On lightcone string field theory
from super Yang-Mills and holography,'' hep-th/0205048.
\bibitem{gmr} D.J. Gross, A. Mikhailov and R. Roiban, ``Operators with large R
charge in N=4 Yang-Mills theory,'' hep-th/0205066.
\bibitem{cfhmm} N.R. Constable, D.Z. Freedman, M. Headrick,
  S. Minwalla, L. Motl, A. Postnikov and W. Skiba, 
``PP-wave string interactions from perturbative Yang-Mills theory,'' 
hep-th/0205089.
\bibitem{sv} M. Spradlin and A. Volovich, ``Superstring interactions in a
pp-wave background,'' hep-th/0204146.
\bibitem{rg} R. Gopakumar, ``String interactions in pp-waves,'' 
hep-th/0205174. 
\bibitem{clp}
M. Cvetic, H. Lu and C.N. Pope, ``Penrose limits, pp-waves and
deformed M2-branes," hep-th/0203082; ``M-theory PP-waves, Penrose
Limits and Supernumerary Supersymmetries,'' hep-th/0203229.
\bibitem{hs} H. Singh, `` M5-branes with 3/8 supersymmetry in pp wave
background,'' hep-th/0205020. 
\bibitem{gns} U. Gursoy, C. Nunez and M. Schvellinger,
``RG flows from Spin(7), CY 4-fold and HK manifolds to AdS, 
Penrose limits and pp waves,'' hep-th/0203124.
\bibitem{jm} J. Michelson, ``(Twisted) toroidal compactification of pp-waves,''
hep-th/0203140.
\bibitem{hks} M. Hatsuda, K. Kamimura and M. Sakaguchi, ``From super-$AdS_5
\times S^5$ algebra to super-pp-wave algebra,'' hep-th/0202190; 
``Super-pp-wave algebra from super-$AdS \times S$ algebras in 
eleven-dimensions,'' hep-th/0204002.
\bibitem{gh} J.P. Gauntlett and C.M. Hull, ``PP-waves in 11-dimensions with
extra supersymmetry,'' hep-th/0203255.
\bibitem{lp} P. Lee and J. Park, ``Open strings in pp-wave background from
defect conformal field theory,'' hep-th/0203257.
\bibitem{vhln} V. Balasubramanian, M. Huang, T.S. Levi and A. Naqvi, ``Open
strings from N=4 super Yang-Mills," hep-th/0204196.
\bibitem{ch1} C. Ahn, ``More on Penrose limit of $AdS_4 \times Q^{1,1,1}$,''
hep-th/0205008; ``Comments on Penrose Limit of $AdS_4 x M^{1,1,1}$,'' 
hep-th/0205109. 
\bibitem{ps} A. Parnachev and D.A. Sahakyan, ``Penrose limit and string
quantization in $AdS_3 \times S^3$,'' hep-th/0205015.
\bibitem{li} M. Li, ``Correspondence principle in a pp-wave background,''
hep-th/0205043.
\bibitem{mss} S.D. Mathur, A. Saxena and Y.K. Srivastava, 
``Scalar propagator in the pp-wave geometry obtained from $AdS_5 \times S^5$,''
hep-th/0205136.
\bibitem{dsr} K. Dasgupta, M.M. Sheikh-Jabbari and M. van Raamsdonk,
``Matrix perturbation theory for M-theory on a pp-wave,''
hep-th/0205185.
\bibitem{gb} G. Bonelli, ``Matrix strings in pp-wave backgrounds from 
deformed super Yang-Mills Theory,''
hep-th/0205213.
\bibitem{kklp} Y. Kiem, Y. Kim, S. Lee, J. Park, ``PP-wave/Yang-Mills
  Correspondence: An Explicit Check,'' hep-th/0205279.
\bibitem{ikm} N. Itzhaki, I. R. Klebanov and S. Mukhi, ``PP Wave Limit 
  and Enhanced Supersymmetry in Gauge Theories,'' hep-th/0202153;
  S. Mukhi, M. Rangamani and E. Verlinde, ``Strings from Quivers,
  Membranes from Moose,'' hep-th/0204174.
\bibitem{go} J. Gomis and H. Ooguri, ``Penrose Limit of ${\cal N}$=1 Gauge 
    Theories,'' hep-th/0202157.
\bibitem{ps} L. A. Pando-Zayas and J. Sonnenschein, ``On Penrose limits
    and gauge theories,'' hep-th/0202186.
\bibitem{as} M. Alishahia and M. M. Sheikh-Jabbari, ``The PP-wave
  Limit of orbifolded $AdS_5\times S^5$,'' hep-th/0203018; ``Strings
  in PP-waves and Worldsheet Deconstruction,'' hep-th/0204174.
\bibitem{kprt} N. Kim, A. Pankiewicz, S-J. Rey and S. Theisen,
  ``Superstring on PP-Wave Orbifold from Large-N Quiver Theory,''
  hep-th/0203080.
\bibitem{dgr} S. R. Das, C. Gomez and S. J. Rey, ``Penrose limits, spontaneous symmetry breaking and holography in pp wave background,'' hep-th/0203164.
\bibitem{lor} R. G. Leigh, K. Okuyama and M. Rozali, ``PP-waves and holography,'' hep-th/0204026.
\bibitem{kp} E. Kiritsis and B. Pioline, ``Strings in homogeneous gravitational waves and null holography,'' hep-th/0204004.
\bibitem{tt} T. Takayanagi and S. Tereshima, ``Strings on Orbifolded
  PP-Waves,'' hep-th/0203093.
\bibitem{ot} K. Oh and R. Tatar, ``Orbifolds, Penrose limits and 
supersymmetry enhancement,'' hep-th/0205067.
\bibitem{rm} R. R. Metsaev, ``Type IIB Green-Schwarz superstring in
  plane wave Ramond-Ramond background,'' Nucl. Phys. B {\bf 625}, 70,
  (2002) hep-th/0112044.
\bibitem{mt} R. R. Metsaev and A. A. Tseytlin, ``Exactly  solvable
  model of superstring in Ramond-Ramond background,'' hep-th/0202109.
\bibitem{rt} J. G. Russo and A. A. Tseytlin, ``On solvable models of 
 type IIB superstrings in NS-NS and R-R plane wave backgrounds,''
 hep-th/0202179.
\bibitem{ft} S. Frolov and A.A. Tseytlin, ``Semiclassical quantization of
rotating superstring in $AdS_5 \times S^5$,'' hep-th/0204226.
\bibitem{dp} A. Dabholakar and S. Parvizi, `` DP Branes in PP-wave
  Background,'' hep-th/0203231.
\bibitem{db} D. Bak, ``Supersymmetric branes in PP Wave Background,''
  hep-th/0204033.
\bibitem{st} K. Skenderis and M. Taylor, `` Branes in AdS and pp-wave
  spacetimes,'' hep-th/0204054. 
\bibitem{bmz} P. Bain, P. Meessen and M. Zamaklar, ``Supergravity
  solutions for D-branes in Hpp-wave background,'' hep-th/0205106.
\bibitem{ak} A. Kumar, R. R. Nayak and  Sanjay, ``D-Brane
  Solutions in pp-wave Background,'' hep-th/0204025; A. Kumar and 
M. Alishahiha,``D-Brane Solutions from New Isometries of pp-Waves,'' hep-th/0205134.
\bibitem{cmt} N. R. Constable, R. C. Myers and O. Tafjord, ``Fuzzy
funnels: Non-abelian Brane Intersections,'' hep-th/0105035.
\bibitem{bp} M. Billo', I. Pesando, ``Boundary States for GS
  superstrings in an Hpp wave background,'' hep-th/0203028.
\bibitem{ch} Chong-Sun Chu and  Pei-Ming Ho, ``Noncommutative D-brane and Open String in pp-wave
Background with B-field,'' hep-th/0203186.   
\bibitem{cgnw} M. Cederwall, A. von Gussich, B. Nilsson and
  A. Westerberg, ``The Dirichlet super-three-brane in ten-dimensional
  type IIB supergravity,'' Nucl. Phys. B {\bf 490}, 163 (1997),
  hep-th/9610148; M. Cederwall, A. von Gussich, P. Sundell and
  A. Westerberg, ``The Dirichlet super-p-branes in ten dimensional
  type IIA and IIB supergravity,'' Nucl. Phys. B {\bf 490}, 179,
  (1997), hep-th/9611159.
\bibitem{aps} M. Aganagic, C. Popescu and J. Schwarz, ``D-brane
  actions with local kappa symmetry,'' Phys. Lett. B {\bf 393}, 311,
  (1997), hep-th/9610249; ``Gauge-invariant and gauge fixed D-brane
  actions,'' Nucl. Phys. B {\bf 495}, 99, (1997), hep-th/9612080.
\bibitem{bt} E. Bergshoeff and P. Townsend, ``Super D-branes,''
  Nucl. Phys. B {\bf 490}, 145, (1997), hep-th/9611173; E. Bergshoeff, 
  R. Kallosh, T. Ortin and G. Papadopoulos, ``Kappa-symmetry,
  supersymmetry and intersecting branes,'' Nucl. Phys. B {\bf 502},
  149, (1997), hep-th/9705040.


  


\end{enumerate}
\end{document}